\documentclass[prl,twocolumn,showpacs,superscriptaddress,nofootinbib]{revtex4-1}
\usepackage[T1]{fontenc} % if needed
\usepackage{amsmath}
\usepackage{amssymb}
\usepackage{amsfonts}
\usepackage{graphicx}
\usepackage{enumitem}
\usepackage{bm}
\usepackage{rotating}
\usepackage{hyperref}
\usepackage{xcolor}
\usepackage{array}
\usepackage{lmodern}
\usepackage[normalem]{ulem}
\usepackage{braket}
\usepackage{tensor}
\usepackage{mathrsfs}
\usepackage{float}

\usepackage[normalem]{ulem}

\begin{document}

\title{Heavy axions from twin dark sectors with $\bar{\theta}$-characterized mirror symmetry}

\author{Pei-Hong Gu}

\email{phgu@seu.edu.cn}

\affiliation{School of Physics, Jiulonghu Campus, Southeast University, Nanjing 211189, China}

\begin{abstract}

The QCD Lagrangian contains a CP violating gluon density term with a physical coefficient $\bar{\theta}$. The upper bound on the electric dipole moment of neutron implies that the value of $\bar{\theta}$ should be extremely small rather than the theoretically expected order of unity. The tiny $\bar{\theta}$ is commonly known as the strong CP problem. In order to solve this puzzle, we construct a $\bar{\theta}$-characterized mirror symmetry between a pair of twin dark sectors with respective discrete symmetries. By taking a proper phase rotation of dark fields, we can perfectly remove the parameter $\bar{\theta}$ from the full Lagrangian. In our scenario, the discrete symmetry breaking, which are responsible for the mass generation of dark colored fermions and dark matter fermions, can be allowed near the TeV scale. This means different phenomena from the popular axion models with high scale Peccei-Quinn global symmetry breaking.

\end{abstract}

%\pacs{14.80.Va, 14.80.Ec, 14.65.Jk, 12.60.Fr}

\maketitle

\textbf{Introduction:} In the standard model (SM), the QCD Lagrangian contains a CP violating gluon density term with a physical coefficient $\bar{\theta}$. The upper bound on the electric dipole moment of neutron enforces that the value of $\bar{\theta}$ should be very tiny rather than the theoretically expected order of unity. The extreme smallness of the value of $\bar{\theta}$ is commonly known as the strong CP problem \cite{kg2010,dgnv2020,sikive2021,workman2022}. 

The most attractive solution to the strong CP problem is the Peccei-Quinn (PQ) global symmetry $U(1)_{\textrm{PQ}}^{}$ proposed in 1977 \cite{pq1977}. After the PQ global symmetry is spontaneously broken, a Goldstone boson can emerge as usual. Subsequently, this Goldstone boson picks up a mass through the color anomaly \cite{adler1969,bj1969,ab1969}. Therefore, the Goldstone boson from the PQ symmetry breaking eventually becomes a pseudo Goldstone boson, named the axion \cite{weinberg1978,wilczek1978}.

The original PQ model within a two Higgs doublet context was quickly ruled out by experimental data. Actually, the axion has not been observed in any experiments so far. This fact implies that the axion should interact with the SM particles at an extremely weak level \cite{kg2010,dgnv2020,sikive2021,workman2022}. To revive the PQ symmetry, Kim-Shifman-Vainstein-Zakharov (KSVZ) \cite{kim1979,svz1980} and Dine-Fischler-Srednicki-Zhitnitsky (DFSZ) \cite{dfs1981,zhitnitsky1980} published their invisible axion models during 1979 to 1981. In the KSVZ-type and DFSZ-type models \cite{kim1979,svz1980,dfs1981,zhitnitsky1980}, a gauge-singlet scalar drives the PQ symmetry to be spontaneously broken far far above the weak scale so that the axion can escape from any experimental limits \cite{kg2010,dgnv2020,sikive2021,workman2022}.

In the invisible axion models, the accompanying new particles should be at the PQ symmetry breaking scale unless the related couplings are artificially small. So, they are too heavy to leave any experimental signals. In other words, all experimental attempts to discover the axion can only depend on the axion-meson mixing \cite{kg2010,dgnv2020,sikive2021,workman2022}. Moreover, for a huge hierarchy between the PQ and electroweak symmetry breaking scales, the inevitable Higgs portal should have an extremely small coupling, otherwise, its contribution should have a large cancellation with the rarely quadratic term of the SM Higgs scalar \cite{cv2016}. In order to realize a low scale PQ symmetry breaking, people have tried to consider an anomalous gauge symmetry \cite{mrs2001} or a huge PQ charge \cite{cky2014,kr2016,ci2016}.

In this paper, we shall consider a pair of twin dark sectors with a $\bar{\theta}$-characterized mirror symmetry in order to solve the strong CP problem without the introduction of PQ symmetry. Specifically, after the dark fermions and scalars take a proper phase rotation, the parameter $\bar{\theta}$ can completely disappear from the full Lagrangian. While two pseudo Goldstone bosons are separately induced by spontaneous breaking of respective discrete symmetries and then are individually coupled to the CP violating gluon density, their zero vacuum expectation values can be guaranteed by the $\bar{\theta}$-characterized mirror symmetry. Our scenario allows that the discrete symmetries can be spontaneously broken near the TeV scale, meanwhile, the induced pseudo Goldstone bosons can keep heavy enough. This means different phenomena from the invisible axion models. Furthermore, the dark fermions include two neutral fermions besides two colored fermions. After the dark fermions obtain their masses through the spontaneous discrete symmetry breaking, the neutral dark fermions can keep stable to account for the dark matter relic density while the colored dark fermions can fast decay into the SM fermions with the dark matter fermions. Remarkably, all of new physics in our scenario can be expected near the TeV scale to verify in various experiments.

\textbf{Strong CP problem:} We briefly review the strong CP problem. The SM QCD Lagrangian is  
\begin{eqnarray}
\label{qcd}
\mathcal{L}_{\textrm{QCD}}^{}&=&\sum_{q}^{} \bar{q} \left(i D \!\!\!\!/ - m_q^{} e^{i\theta_q}_{} \right)q - \frac{1}{4} G^{a\mu\nu}_{} G_{\mu\nu}^{a} \nonumber\\
&& - \theta \frac{\alpha_s^{}}{8\pi} G^{a\mu\nu}_{}\tilde{G}_{\mu\nu}^{a}\,.
\end{eqnarray}
Here $\theta_q^{}$ is the phase from the Yukawa couplings of quark fields, $\theta$ is the QCD vacuum angle, $\alpha_s^{}$ is the strong coupling constant, $G^{a}_{\mu\nu}$ is the gluon field strength tensor and $\tilde{G}^{a}_{\mu\nu}$ is its dual. Through the chiral phase transformation of quark fields, i.e.
\begin{eqnarray}
q\rightarrow e^{-i \gamma_5^{} \theta_q/2}_{}  q \,,
\end{eqnarray}
the phases of quark mass terms can be removed from the QCD Lagrangian, i.e. 
\begin{eqnarray}
\label{qcdphy}
\mathcal{L}_{\textrm{QCD}}^{}&=&\sum_{q}^{} \bar{q} \left(i D \!\!\!\!/ - m_q^{} \right)q - \frac{1}{4} G^{a\mu\nu}_{} G_{\mu\nu}^{a}  - \bar{\theta} \frac{\alpha_s^{}}{8\pi} G^{a\mu\nu}_{}\tilde{G}_{\mu\nu}^{a}\nonumber\\
&&\textrm{with}~~\bar{\theta}\equiv\theta- \textrm{Arg}\textrm{ Det} \left(M_d^{} M_u^{} \right)\,.
\end{eqnarray}
In the above, $M_{d}^{}$ and $M_{u}^{}$ are the respective mass matrices of the SM down-type and up-type quarks. The upper limits on the electric dipole moment of neutron implies that the physical parameter $\bar{\theta}$ should have an extremely small value rather than the theoretically expected order of unity, i.e.
\begin{eqnarray}
\left|\bar{\theta}\right|<10^{-10}_{}\,.
\end{eqnarray}
This fine tuning of ten orders of magnitude is commonly known as the strong CP problem.

\begin{table*}
%\vspace{0.25cm}
\begin{center}
\begin{tabular}{|l||c|c|c|c|c||c|c|c|c|c||c|}  \hline &&&&&&&&&&&\\[-2.0mm] ~~~$Scalars~\&~Fermions$~~~&~~~$\xi_1^{}$~~~&~~~$\psi_{L1}^{}$~~~&~~~$\psi_{R1}^{}$~~~&~~~$\chi_{L1}^{}$~~~&~~~$\chi_{R1}^{}$~~~& ~~~$\xi_{2}^{}$~~~&~~~$\psi_{L2}^{}$~~~&~~~$\psi_{R2}^{}$~~~&~~~$\chi_{L2}^{}$~~~&~~~$\chi_{R2}^{}$~~~&~~~$\omega$~~~ \\
&&&&&&&&&&&\\[-2.0mm]\hline&&&&&&&&&&& \\[-1.5mm]~~~~~~~~~~~~spin&$0$ &$\frac{1}{2}$ &$\frac{1}{2}$ &$\frac{1}{2}$ &$\frac{1}{2}$ &$0$ & $\frac{1}{2}$ & $\frac{1}{2}$&$\frac{1}{2}$ &$\frac{1}{2}$& $0$ \\
&&&&&&&&&&&\\[-2.0mm]\hline&&&&&&&&&&& \\[-1.5mm]~~~~~~~~~~~~color&$1$ &$3$ &$3$ &$1$ &$1$ &$1$ & $3$ & $3$&$1$ &$1$ &$3$ \\
&&&&&&&&&&&\\[-2.0mm]\hline&&&&&&&&&&& \\[-1.5mm]~~~~~~~~~~~~hypecharge&$0$ &$-\frac{1}{3}$ &$-\frac{1}{3}$ &$0$ &$0$ &$0$ & $-\frac{1}{3}$ & $-\frac{1}{3}$&$0$ &$0$ &$-\frac{1}{3}$ \\
&&&&&&&&&&&\\[-2.0mm]\hline&&&&&&&&&&& \\[-1.5mm]~~~~~~~~~~~~$Z_{8}^{(1)}$&$i$ &$e^{+i\pi/4}_{}$ &$e^{-i\pi/4}_{}$ & $e^{-i\pi/4}_{}$& $e^{+i\pi/4}_{}$ &$1$ & $1$ & $1$&$1$ &$1$ &$1$  \\
&&&&&&&&&&&\\[-2.0mm]\hline&&&&&&&&&&& \\[-1.5mm]~~~~~~~~~~~~$Z_{8}^{(2)}$ &$1$&$1$  & $1$& $1$ &$1$ &  $i$ & $e^{+i\pi/4}_{}$&$e^{-i\pi/4}_{}$ &$e^{-i\pi/4}_{}$ &$e^{+i\pi/4}_{}$ &$1$ \\
[1.5mm]
\hline
\end{tabular}
\vspace{0.25cm}
\caption{\label{fields} The non-SM scalars and fermions. The fields with indices "1" and "2" form a pair of twin dark sectors while the colored scalar $\omega$ is a messenger between the dark sectors and the SM.}
\end{center}
\end{table*}

\textbf{Twin dark sectors:} The non-SM scalars and fermions are summarized in Table. \ref{fields}, where the fields with indices "1" and "2" form a pair of twin dark sectors while the colored scalar $\omega$ is a messenger between the dark sectors and the SM. We further impose a $\bar{\theta}$-characterized mirror symmetry, under which the dark sector $1$ and the dark sector $2$ transform as 
\begin{eqnarray}
\label{mirror}
\!\!\!\!\!\!\!\!\left(\begin{array}{l}
\xi_{1}^{} \\
[3mm]
\psi_{L1}^{}\\
[3mm]
\psi_{R1}^{}\\
[3mm]
\chi_{L1}^{}\\
[3mm]
\chi_{R1}^{}
\end{array}\right)& \!\!\stackrel{\textrm{~$\bar{\theta}$-characterized mirror~symmetry }}{\leftarrow\!\!\!-\!\!\!-\!\!\!-\!\!\!-\!\!\!-\!\!\!-\!\!\!-\!\!\!-\!\!\!-\!\!\!-\!\!\!-\!\!\!-\!\!\!-\!\!\!-\!\!\!-\!\!\!-\!\!\!-\!\!\!-\!\!\!-\!\!\!-\!\!\!-\!\!\!\rightarrow} \!\!& \left(\begin{array}{l}
e^{-i\bar{\theta}}_{}\xi_{2}^{} \\
[3mm]
e^{-i\bar{\theta}/2}_{}\psi_{L2}^{}\\
[3mm]
e^{+i\bar{\theta}/2}_{}\psi_{R2}^{}\\
[3mm]
e^{+i\bar{\theta}/2}_{}\chi_{L2}^{}\\
[3mm]
e^{-i\bar{\theta}/2}_{}\chi_{R2}^{}\end{array}\right). \end{eqnarray}

For simplicity, we do not write down the full Lagrangian, which is constrained at renormalizable level. Because of our chosen charge assignments in Table \ref{fields}, all of the allowed Yukawa interactions involving the non-SM fields are
\begin{eqnarray}
\label{yukawa}
\mathcal{L}_Y^{}&\supset& -~ y_\psi^{}\left[\left(\xi_{1}^{} \bar{\psi}_{L1}^{}  \psi_{R1}^{} + \xi_{2}^{} \bar{\psi}_{L2}^{} \psi_{R2}^{}\right)+\textrm{H.c.}\right]\nonumber\\
[2mm]
&& - ~y_\chi^{} \left[\left(\xi_{1}^{}\bar{\chi}_{R1}^{} \chi_{L1}^{} + \xi_{2}^{} \bar{\chi}_{R2}^{}  \chi_{L2}^{}\right)+ \textrm{H.c.}\right]\nonumber\\
[2mm]
&&-~y^{}_{LR}\left[\omega\left( \bar{\psi}_{L1}^{} \chi_{R1}^{} + \bar{\psi}_{L2}^{} \chi_{R2}^{} \right)+\textrm{H.c.}\right] \nonumber\\
[2mm]
&&-~y^{}_{RL}\left[\omega\left( \bar{\psi}_{R1}^{} \chi_{L1}^{} + \bar{\psi}_{R2}^{} \chi_{L2}^{} \right)+\textrm{H.c.}\right]\nonumber\\
[2mm]
&&+~\textrm{the~couplings~of}~\omega~\textrm{to~the~SM~fermions}\,,
\end{eqnarray}
meanwhile, the full potential of the dark Higgs scalars is
\begin{eqnarray}
\label{potential}
V(\xi_1^{},\xi_2^{})&=& \mu_\xi^2 \left(\xi_1^\ast \xi_1^{}+\xi_2^\ast\xi_2^{}\right) +\lambda_\xi^{}\left[\left(\xi_1^\ast \xi_1^{}\right)^2_{} + \left(\xi_2^\ast \xi_2^{}\right)^2_{}\right] \nonumber\\
[2mm]
&&+\lambda'^{}_\xi  \xi_1^\ast \xi_1^{} \xi_2^\ast\xi_2^{}+\kappa_\xi^{} \left[\left(\xi_1^4+e^{-i4 \bar{\theta}}_{}\xi_2^4 \right)+\textrm{H.c.}\right]\,.\nonumber\\
&&
\end{eqnarray}
We would like to emphasize that the full Lagrangian exactly respects the $\bar{\theta}$-characterized mirror symmetry (\ref{mirror}).

It is easy to check while the other terms in the full Lagrangian can keep invariant, the SM QCD Lagrangian (\ref{qcdphy}) and the dark scalar potential (\ref{potential}) can simultaneously get rid of the parameter $\bar{\theta}$, i.e. 
\begin{eqnarray}
\label{phyqcd}
\mathcal{L}_{\textrm{QCD}}^{}& \Rightarrow &\sum_{q}^{} \bar{q} \left(i D \!\!\!\!/ - m_q^{} \right)q - \frac{1}{4} G^{a\mu\nu}_{} G_{\mu\nu}^{a}\,,\\
[3mm]
\label{phypotential}
V(\xi_1^{},\xi_2^{})&\Rightarrow& \mu_\xi^2 \left(\xi_1^\ast \xi_1^{}+\xi_2^\ast\xi_2^{}\right) +\lambda_\xi^{}\left[\left(\xi_1^\ast \xi_1^{}\right)^2_{} + \left(\xi_2^\ast \xi_2^{}\right)^2_{}\right] \nonumber\\
[2mm]
&&+\lambda'^{}_\xi  \xi_1^\ast \xi_1^{} \xi_2^\ast\xi_2^{}+\kappa_\xi^{} \left[\left(\xi_1^4+\xi_2^4 \right)+\textrm{H.c.}\right]\,,
\end{eqnarray}
after the dark fields take the phase rotation as below,
\begin{eqnarray}
\left(\begin{array}{l}\psi_{L1}^{}\rightarrow  \psi_{L1}\\
[3mm] \psi_{R1}^{}\rightarrow  \psi_{R1}\\
[3mm]
\chi_{L1}^{}\rightarrow  \chi_{L1}\\
[3mm]\chi_{R1}^{}\rightarrow  \chi_{R1} \\
[3mm]
\psi_{L2}^{} \rightarrow e^{+i\bar{\theta}/2}_{} \psi_{L2}^{}\\
[3mm]
\psi_{R2}^{} \rightarrow e^{-i\bar{\theta}/2}_{} \psi_{R2}^{}\\
[3mm]
\chi_{L2}^{} \rightarrow e^{-i\bar{\theta}/2}_{} \chi_{L2}^{}\\
[3mm]
\chi_{R2}^{} \rightarrow e^{+i\bar{\theta}/2}_{} \chi_{R2}^{}\end{array}\right) \Longrightarrow \left(\begin{array}{l}\xi_{1}^{}\rightarrow   \xi_{1} \\
[3mm] \xi_{2}^{} \rightarrow e^{+i\bar{\theta}}_{} \xi_2^{}\end{array} \right)\,.
\end{eqnarray}
This means the twin dark sectors in Table \ref{fields} with the $\bar{\theta}$-characterized mirror symmetry in Eq. (\ref{mirror}) certainly can help us to realize a physical Lagrangian without the unexpected parameter $\bar{\theta}$.

\textbf{Heavy axions:} When the dark Higgs scalars $\xi_{1,2}^{}$ develop their nonzero vacuum expectation values $v_{1,2}^{}$, they can be expressed by  
\begin{eqnarray}
\xi_1^{} =\frac{1}{\sqrt{2}}\left(v_{1}^{}+h_{1}^{}\right) e^{ia_1^{}/v_{1}^{}}_{}\,,~~\xi_2^{} = \frac{1}{\sqrt{2}}\left(v_{2}^{}+h_{2}^{}\right) e^{ia_2^{}/v_{2}^{}}_{}\,,\nonumber\\
&&
\end{eqnarray}
with $h_{1,2}^{}$ being the Higgs bosons and $a_{1,2}^{}$ being the Goldstone bosons. Due to the $\kappa_\xi^{}$-term in the physical scalar potential (\ref{phypotential}), the Goldstone bosons $a_{1,2}^{}$ acquire the following potential and hence become the pseudo Goldstone bosons, i.e.
\begin{eqnarray}
V(a_1^{}, a_2^{})&=&  \frac{1}{2}\kappa_\xi^{}\left[v_1^4 \cos \left(4a_1^{}/v_1^{}\right) + v_2^4 \cos \left(4a_2^{}/v_2^{}\right) \right]\nonumber\\
[2mm]
&\approx&\frac{1}{2}m_{1}^2 a_1^2 + \frac{1}{2}m_2^2 a_2^2 \quad \textrm{with} ~~m_{1,2}^2\equiv -8 \kappa_\xi^{} v_{1,2}^2\,.\nonumber\\
&&
\end{eqnarray}
Here we have expanded the cosine functions for $a_{1,2}^{}/v_{1,2}^{}\ll 1$ by neglecting an irrelevant constant. Clearly, the above potential arrives at its minimum for the zero vacuum expectation values, i.e.
\begin{eqnarray}
\label{vevpgb}
&&\frac{\partial V(a_1^{},a_2^{})}{\partial \langle a_1^{}\rangle} = \frac{\partial V(a_1^{},a_2^{})}{\partial \langle a_2^{}\rangle} =0~~\textrm{for}~~\langle a_{1}^{}\rangle =0\,,
\langle a_{2}^{}\rangle =0\,.\nonumber\\
&&
\end{eqnarray}

Through their Yukawa couplings to the dark Higgs scalars $\xi_{1,2}^{}$, the colored dark fermions $\psi_{1,2}^{}$ acquire the mass terms as follows, 
\begin{eqnarray}
\label{masspsi}
\mathcal{L}&\supset& -m_{\psi_1}^{}\bar{\psi}_{L1}^{} \psi_{R1}^{} e^{i a_{1}^{}/v_{1}^{}}_{} -m_{\psi_2}^{}\bar{\psi}_{L2}^{} \psi_{R2}^{} e^{i a_{2}^{}/v_{2}^{}}_{}+\textrm{H.c.}\nonumber\\
[2mm]
&&\textrm{with}~~m_{\psi_{1,2}}^{}\equiv -\frac{1}{\sqrt{2}}y_\psi^{}v_{1,2}^{}\,.
\end{eqnarray}
We then remove the pseudo Goldstone bosons $a_{1,2}^{}$ from the above mass terms by rephasing the colored dark fermions $\psi_{1,2}^{}$, i.e.
\begin{eqnarray}
\!\!\!\!\!\!\!\left\{\begin{array}{l}\psi_{L1}^{} \rightarrow e^{+i a_{1}^{}/ 2v_{1}^{}}_{}\psi_{L1}^{}\,,\\
[3mm]
\psi_{R1}^{} \rightarrow e^{-i a_{1}^{}/ 2v_{1}^{}}_{}\psi_{R1}^{}\,;\end{array}\right.~~
\left\{\begin{array}{l}\psi_{L2}^{} \rightarrow e^{+i a_{2}^{}/ 2v_{2}^{}}_{}\psi_{L2}^{}\,,\\
[3mm]
\psi_{R2}^{} \rightarrow e^{-i a_{2}^{}/ 2v_{2}^{}}_{}\psi_{R2}^{}\,.\end{array}\right.
\end{eqnarray}
As a result, the pseudo Goldstone bosons $a_{1,2}^{}$ appear in the kinetic terms of the colored dark fermions $\psi_{1,2}^{}$, i.e.
\begin{eqnarray}
\label{axial}
\mathcal{L}&\supset& \frac{1}{2v_{1}^{}} \left(\partial_\mu^{}a_{1}^{} \right)\bar{\psi}_{1}^{} \gamma^\mu_{}  \gamma_5^{}\psi_{1}^{} +\frac{1}{2v_{2}^{}} \left(\partial_\mu^{}a_{2}^{}\right)\bar{\psi}_1^{} \gamma^\mu_{} \gamma_5^{}\psi_2^{}\nonumber\\
[2mm]
&=& -\frac{a_1^{}}{2v_{1}^{}}\, \partial_\mu^{}\!\left(\bar{\psi}_1^{} \gamma^\mu_{} \gamma_5^{} \psi_1^{} \right)- \frac{a_{2}^{}}{2v_{2}^{}} \,\partial_\mu^{}\!\left(\bar{\psi}_2^{} \gamma^\mu_{} \gamma_5^{}\psi_2^{}\right)\nonumber\\
[2mm]
&=& -\left(\frac{a_{1}^{}}{v_{1}^{}} +\frac{a_{2}^{}}{v_{2}^{}} \right)\!\!\left(\frac{\alpha}{72\pi}F_{\mu\nu}^{} \tilde{F}^{\mu\nu}_{}+\frac{\alpha_s^{}}{8\pi}G_{\mu\nu}^{a} \tilde{G}^{a\mu\nu}_{} \right)\nonumber\\
[2mm]
&&-i\frac{m_{\psi_1}^{}}{v_1^{}}a_1^{} \bar{\psi}_{1}^{} \gamma_5^{} \psi_{1}^{}-i\frac{m_{\psi_2}^{}}{v_2^{}}a_2^{} \bar{\psi}_{2}^{} \gamma_5^{} \psi_{2}^{}\,,
\end{eqnarray}
where $\alpha$ is the fine-structure constant, $F_{\mu\nu}^{}$ is the photon filed strength and $\tilde{F}_{\mu\nu}^{}$ is for the dual, while $\alpha_s^{}$, $G^{a}_{\mu\nu}$ and $\tilde{G}^{a}_{\mu\nu}$ have been previously clarified below Eq. (\ref{qcd}).

We now can conclude that the pseudo Goldstone bosons $a_{1,2}^{}$ play the role of axion. This is because they are endowed with the zero vacuum expectation values and are coupled to the CP violating gluon density, as shown in Eqs. (\ref{vevpgb}) and (\ref{axial}). Of course, the pseudo Goldstone bosons $a_{1,2}^{}$ have different properties compared with the conventionally ultralight axions from high scale PQ global symmetry breaking. In particular, both their masses $m_{a_{1,2}}^{}$ and their decay constants $v_{1,2}^{}$ can be allowed near the TeV scale. 

We may recall that the invisible axion models always suffer a so-called PQ quality problem \cite{gh1981,ds1986,bs1992,kmr1992,kmr1992,hhkkww1992,glr1992}. This is because quantum gravity effects must violate all global symmetries at some level \cite{hawking1975}. Such general consensus implies that the PQ global symmetry should be explicitly broken down to a discrete symmetry by some higher-order nonrenormalizable operators. The parameter $\bar{\theta}$ can eventually appear in these operators after it is removed from the SM QCD Lagrangian. The invisible axion then acquires a nonzero vacuum expectation value depending on the parameter $\bar{\theta}$. Unless this vacuum expectation value is small enough, it will destroy the effort of the PQ symmetry to solve the strong CP problem. In order to moderate the PQ quality problem, several mechanisms have been put forth \cite{dgnv2020}.

\textbf{Dark matter:} The neutral dark fermions $\chi_{1,2}^{}$ acquire the following masses through their Yukawa couplings with the dark Higgs scalars $\xi_{1,2}^{}$, i.e.
\begin{eqnarray}
\label{masspsi}
\mathcal{L}&\supset& -m_{\chi_1}^{}\bar{\psi}_{R1}^{} \chi_{L1}^{} e^{i a_{1}^{}/v_{1}^{}}_{} -m_{\chi_2}^{}\bar{\psi}_{R2}^{} \psi_{L2}^{} e^{i a_{2}^{}/v_{2}^{}}_{}+\textrm{H.c.}\nonumber\\
[2mm]
&&\textrm{with}~~m_{\chi_{1,2}}^{}\equiv -\frac{1}{\sqrt{2}}y_\chi^{}v_{1,2}^{}\,.
\end{eqnarray} 
The pseudo Goldstone bosons $a_{1,2}^{}$ can be removed from the above mass terms after the neutral dark fermions  $\chi_{1,2}^{}$ take the phase rotation as below, 
\begin{eqnarray}
\!\!\!\!\!\!\!\left\{\begin{array}{l}\chi_{L1}^{} \rightarrow e^{-i a_{1}^{}/ 2v_{1}^{}}_{}\chi_{L1}^{}\,,\\
[3mm]
\chi_{R1}^{} \rightarrow e^{+i a_{1}^{}/ 2v_{1}^{}}_{}\chi_{R1}^{}\,;\end{array}\right.~~
\left\{\begin{array}{l}\chi_{L2}^{} \rightarrow e^{-i a_{2}^{}/ 2v_{2}^{}}_{}\chi_{L2}^{}\,,\\
[3mm]
\chi_{R2}^{} \rightarrow e^{+i a_{2}^{}/ 2v_{2}^{}}_{}\chi_{R2}^{}\,.\end{array}\right.
\end{eqnarray}
Consequently, the pseudo Goldstone bosons $a_{1,2}^{}$ are devolved to the kinetic terms of the neutral dark fermions $\chi_{1,2}^{}$, i.e.
\begin{eqnarray}
\mathcal{L}&\supset&- \frac{1}{v_{1}^{}} \left(\partial_\mu^{}a_{1}^{} \right)\bar{\chi}_{1}^{} \gamma^\mu_{}  \gamma_5^{}\chi_{1}^{} - \frac{1}{v_{2}^{}} \left(\partial_\mu^{}a_{2}^{}\right)\bar{\chi}_2^{} \gamma^\mu_{} \gamma_5^{}\chi_2^{}\nonumber\\
[2mm]
&=& \frac{a_1^{}}{v_{1}^{}}\, \partial_\mu^{}\!\left(\bar{\chi}_1^{} \gamma^\mu_{} \gamma_5^{} \chi_1^{} \right)+ \frac{a_{2}^{}}{v_{2}^{}} \,\partial_\mu^{}\!\left(\bar{\chi}_2^{} \gamma^\mu_{} \gamma_5^{}\chi_2^{}\right)\nonumber\\
[2mm]
&=&i\frac{m_{\chi_1}^{}}{v_1^{}}a_1^{} \bar{\chi}_{1}^{} \gamma_5^{} \chi_{1}^{}+i\frac{m_{\chi_2}^{}}{v_2^{}}a_2^{} \bar{\chi}_{2}^{} \gamma_5^{} \chi_{2}^{}\,.
\end{eqnarray}

The colored dark fermions $\psi_{1,2}^{}$ are assumed heavy enough so that one colored dark fermion $\psi_{1,2}^{}$ can fast decay into one real or virtual colored messenger scalar $\omega$ with one neutral dark fermion $\chi_{1,2}^{}$, subsequently, one colored messenger scalar $\omega$ can fast decay into two SM quarks and/or one SM quark and one SM lepton. Therefore, the neutral dark fermions $\chi_{1,2}^{}$ can keep stable to leave a relic density in the present universe.

The stable fermions $\chi_{1,2}^{}$ can have the $t$-channel annihilations into two Higgs bosons $h_{1,2}^{}$, two pseudo Goldstone bosons $a_{1,2}$, and/or one Higgs boson $h_{1,2}^{}$ and one pseudo Goldstone boson $a_{1,2}^{}$, as long as the kinematics is allowed. In addition, the stable fermions $\chi_{1,2}^{}$ can have the $s$-channel annihilations into the SM fields through the Higgs portal between the dark and SM Higgs scalars. With these annihilations, the stable fermions $\chi_{1,2}^{}$ can contribute a right relic density to serve as the dark matter particles. Meanwhile, the dark matter fermions $\chi_{1,2}^{}$ can scatter off nuclei due to the dark and SM Higgs mixing.

\textbf{Conclusions and discussions:} In the present paper, we have proposed a new mechanism to solve the strong CP problem by introducing a $\bar{\theta}$-characterized mirror symmetry between a pair of twin dark sectors with respective discrete symmetries. Specifically, after the dark fermions and scalars take a proper phase rotation, the parameter $\bar{\theta}$ can completely disappear from the full Lagrangian. Our scenario can allow that the discrete symmetries of the dark fields are spontaneously broken near the TeV scale, meanwhile, the induced pseudo Goldstone bosons keep heavy enough to satisfy any experimental constraints. This is very different from the conventional axion models with high sale PQ global symmetry breaking. After the dark fermions obtain their masses through the spontaneous discrete symmetry breaking, the colored dark fermions can fast decay into the SM fermions with the neutral dark fermions, while the neutral dark fermions can keep stable to account for the dark matter relic density. In our scenario, the new particles are not required far above the TeV scale so that all of them can be expected to verify in various experiments.

The $\bar{\theta}$-characterized mirror symmetry predicts the degenerate mass spectra of the twin dark fields. This degeneracy may be unfortunately excluded in experiments. In this case, we can revive our mechanism in a slightly extended model where a real singlet scalar carries an odd transformation under the $\bar{\theta}$-characterized mirror symmetry. This treatment is just like the spontaneous D-parity violation in some left-right symmetric models \cite{cm1987}.

\textbf{Acknowledgement}: This work was supported in part by the National Natural Science Foundation of China under Grant No. 12175038.

\end{document}